\documentclass[useAMS,usegraphicx,usenatbib]{mn2e}
\usepackage{amsmath}

\title[Evidence that widespread star formation may be underway in G0.253+0.016, ``The Brick"]{Evidence that widespread star formation may be underway in G0.253+0.016, ``The Brick"}
\author[K. A. Marsh et al.]{K. A. Marsh$^{1}$\thanks{E-mail:
Ken.Marsh@astro.cf.ac.uk}, 
S. E. Ragan$^{2}$,
A. P. Whitworth$^{1}$, 
\& P. C. Clark$^{1}$\\
$^{1}$School of Physics and Astronomy, Cardiff University, Cardiff CF24 3AA, UK\\
$^{2}$School of Physics and Astronomy, The University of Leeds, Leeds, LS2 9JT, UK}
\begin{document}

\pagerange{\pageref{firstpage}--\pageref{lastpage}} \pubyear{2016}

\maketitle

\label{firstpage}

\begin{abstract}
Image cubes of differential column density as a function of dust temperature
are constructed for Galactic Centre molecular cloud G0.253+0.016 
(``The Brick") using the recently described {\tt PPMAP} procedure. The input 
data consist of continuum images from the {\it Herschel\/} 
Space Telescope in the wavelength range 70--500 $\mu$m, supplemented
by previously published interferometric data at 1.3 mm wavelength.
While the bulk of the dust in the molecular cloud is consistent with being
heated externally by the local interstellar radiation field, our image cube
shows the presence, near one edge of the cloud, of a filamentary structure whose
temperature profile suggests internal heating. The structure appears as a cool 
($\sim14$ K) tadpole-like feature, $\sim6$ pc in length, in which is 
embedded a thin spine of much hotter ($\sim40$--50 K) material. We interpret 
these findings in terms of a cool filament whose hot central region is 
undergoing gravitational collapse and
fragmentation to form a line of protostars.
If confirmed, this would represent the first evidence 
of widespread star formation having started within this cloud.
\end{abstract}


\section{Introduction}
The dense molecular cloud, G0.253+0.016 (also known as ``The Brick"), has been
the subject of intense recent study as
a likely future site of prolific high-mass star formation \citep{long12,rath14}.
However, with the exception of a 22\,GHz water maser 
source \citep{lis94}, likely associated with a self gravitating core
\citep{rath14b}, no clear evidence of star formation has been 
reported to date. More recent observations at centimetre wavelengths 
\citep{mills15} have not revealed any additional maser activity attributable 
to star formation. 
Thus the question remains: what has delayed or is preventing prodigious 
star formation in the Brick? Dynamical arguments, 
such as its highly turbulent nature \citep{kauff13,john14,rath15} or its 
apparent location within a coherent ring-like stream of molecular clouds 
in orbit about Sgr A$^*$ \citep{mol2011,long13,kruij15} 
have been cited as possible agents of star formation suppression. 
In order to set more stringent constraints on the level of star formation
activity, it would seem worthwhile to examine other possible indicators, such
as the presence of localised peaks in dust temperature that might signal
the presence of protostars at early stages of evolution. With this in mind,
we present here the results of a detailed study of the variation of dust 
temperature within the cloud.

Studies of the temperature structure of the Brick show that the
gas temperature is uniformly higher than the dust temperature and that
those two components are heated by different mechanisms. The observed
dust temperatures of $T_{\rm D}\sim 19$--27 K \citep{long12}  are consistent 
with heating by the
local interstellar radiation field (ISRF) assuming that the latter is $\sim1000$
times stronger than that in the solar vicinity \citep{clark13}. However,
the gas \citep[$T_{\rm G}\sim60$--100 K]{guesten81,mills13} is most likely 
heated by a combination of cosmic rays \citep{clark13} and turbulent 
compression \citep{gin16}, with the former dominating at high densities.

Our study of the dust temperature distribution
makes use of the {\tt PPMAP} procedure \citep{mar15}, whose purpose is to
generate resolution-enhanced image cubes of
differential column density as a function of dust temperature, $T_{\rm D}$,
and sky position ($x,y$),
given sets of observed images of dust continuum emission at
multiple wavelengths.  Differential column density, in this case, is 
defined as the column density of material (gas plus dust) per unit 
interval of dust temperature, and is expressed in units of
hydrogen molecules per square centimetre per degree Kelvin.
{\tt PPMAP} is a Bayesian technique whose output is
the expectation value of the differential column density in each cell of
an image cube whose axes are ($x,y,T_{\rm D}$), given the input data.
The {\it a priori\/} probability of cell occupancy is controlled by
a ``dilution" parameter, $\eta$, such that the smaller the value of $\eta$,
the more the algorithm tries to fit the data with as few source
components as possible. The exact value of $\eta$ is not critical, but
the most appropriate value is one which results in
a reduced chi squared value of order unity, indicating that the data
have been fit within the error bars using a minimum number of source
components. The procedure also yields a corresponding image cube of
uncertainty values. Errors are Poisson-like, and increase in regions of
high total column density.
A key underlying assumption is that the dust is optically thin 
at all wavelengths of observation. {\tt PPMAP} takes full account of the point 
spread functions (PSFs) of the telescopes used, and does not require all 
images to be degraded to the same resolution. 

\section[]{Observational data and analysis procedure}

The principal inputs to {\tt PPMAP} consist of a set of observed images of 
dust continuum emission, their associated PSFs, the assumed dust opacity 
law over the wavelength range of the observations, and a grid of temperature 
values at which the differential column density will be estimated. The output 
is then an image cube of differential column density as a function of 
angular position on the set of predefined temperature planes, plus a
corresponding image cube of uncertainty values.

Our observational data consist of a set of {\em Herschel\/} PACS and SPIRE
images at wavelengths of 70 $\mu$m\footnote{Although the Brick is known to
be optically thick ($\tau\sim1$) at 70 $\mu$m (see, for example,
\citealt{long12}), we include the 70 $\mu$m image in order to make use of the
information it provides on the possible presence of hot material,
of particular value when searching for potential sites of star formation.
Although this violates a key assumption in {\tt PPMAP}, the inclusion of 
a single wavelength for which the source is partially optically thick does 
not seriously impact the temperature and column density estimates, since those 
quantities are well constrained by the longer-wavelength data within the
optically-thin regime. The benefit is that it
facilitates the detection of any hot components
whose column densities may be too low for detection at longer wavelengths.
Tests with synthetic data confirm these expectations.},
160 $\mu$m, 250 $\mu$m, 350 $\mu$m, and 
500 $\mu$m, observed as part of the Hi-GAL survey \citep{mol2010}. 
The spatial resolution values of these maps, i.e., the beam sizes at full 
width half maximum (FWHM), are approximately
$8.5''$, $13.5''$, $18.2''$, $24.9''$, and $36.3''$, respectively.
The associated PSFs are 
based on the measured {\it Herschel\/} beam profiles \citep{pog10,griffin13}. 
The adopted temperature grid consists of 12 temperatures equally spaced 
in $\log T_{\rm D}$ between 7 K and 60 K. These temperatures may be regarded
as the approximate midpoints of a set of finite intervals, the $i^{\rm th}$ of
which may be expressed as $[(T_{i\!-\!1}\!+\!T_i)/2,\,(T_i\!+\!T_{i\!+\!1})/2]$.
The spatial sampling interval of the output grid is $4''$ per pixel.  

Following \citet{rath15}, we adopt the following form for the variation
of dust opacity with wavelength:
\begin{equation}
\kappa(\lambda) = 0.042\,{\rm cm}^2\,{\rm g}^{-1}\,\left(\frac{\lambda}{300\,\mu{\rm m}}\right)^{-1.2}\,.
\label{eq12}
\end{equation}
in which the power law index is based on the \citet{rath15} estimate of
$\beta=1.2\pm0.1$, and the coefficient is extrapolated from their 1.2 mm
value of 0.8 cm$^2$ g$^{-1}$ based on \citet{oss94}.
It is expressed in terms of the total mass (dust plus gas).
In comparing the observed fluxes with model values, allowance must be
made for the fact the observations represent averages over finite
bandpasses rather than monochromatic values. Since the published PACS and 
SPIRE fluxes are based on an assumed source
spectrum which is flat in $\nu F_\nu$, allowance must therefore be made for 
other spectral shapes. In {\tt PPMAP}, temperature-dependent
colour corrections are applied to the model images using the tables 
presented by \citet{pez13} and \citet{valt14}.

\begin{figure*}
\includegraphics[width=120mm]{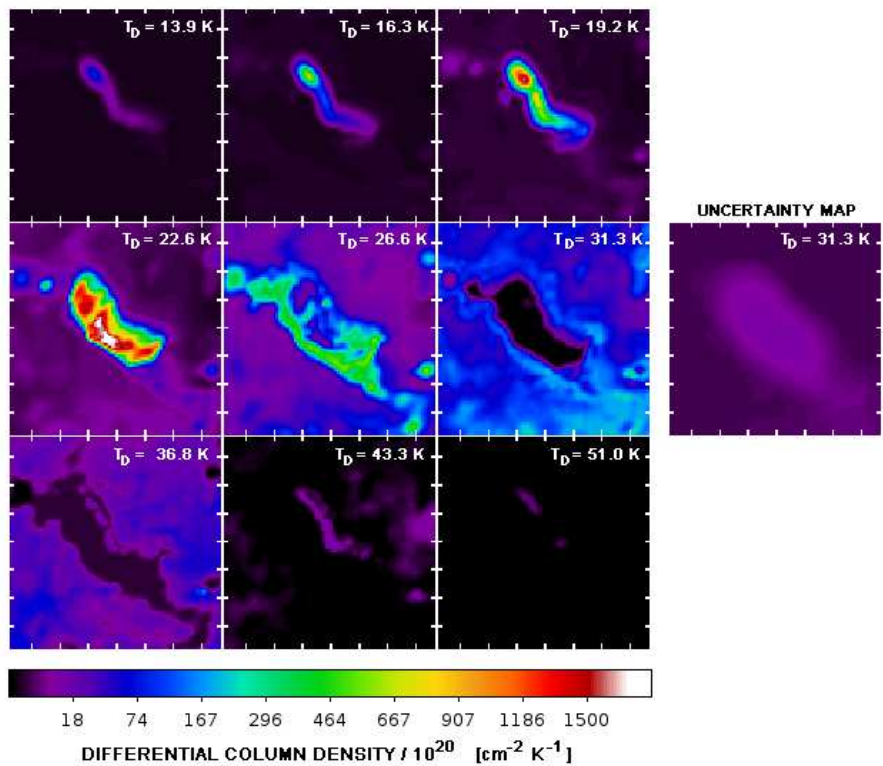}
\caption{Estimated differential column-density of the Brick in nine 
different planes of dust temperature, $T_{\rm D}$,
as indicated in the top right of each panel. The differential
column density is expressed in terms of the number density of H$_2$
molecules. The accompanying uncertainty map for the $T_{\rm D}=31.3$ K case  
illustrates that even when the map shows a deep depression there 
may still exist a substantial amount of material 
below the sensitivity limit. The field of view
in each case is $12.8' \times 12.8'$ and the pixel size is $4''$ 
(0.16 pc at the assumed 8.4 kpc distance). The tick marks are at
intervals of 2 pc.  The images were generated by 
applying {\tt PPMAP} using {\em Herschel\/} data in the wavelength range 
70--500 $\mu$m.}
\label{fig1}
\end{figure*}

Fig. \ref{fig1} shows the results in the form of a set
of differential column density images for each of the nine dust temperature
planes in which significant column density is detected.
We have also produced an image cube of higher spatial resolution
by supplementing the {\em Herschel\/} data with an interferometric 
image at 1.3 mm wavelength, published by \citet{john14}, based on observations
made with the Submillimeter Array (SMA). In doing so,
the SMA image was treated in the same way as for the {\it Herschel\/}
images, i.e., as a separate input image with its own PSF. The properties
of the latter were determined by the interferometer configuration of the
SMA whereby the lack of baselines shorter than 16.4 m had the effect of
suppressing spatial information on angular scales $\stackrel{>}{_\sim}20''$,
leading to a zero-sum PSF. Also, since the resolution of the 1.3 mm
image ($4.3'' \times 2.7''$) exceeds the Nyquist limit for
representation on the $4''$ output grid, the 1.3 mm image was 
smoothed by a Gaussian designed to produce an effective beam FWHM 
of $8''$, and the same smoothing was applied to the PSF. The result
is shown in Fig.\ 1.1 which, due to space limitations, is available
in the on-line version of this Letter only.

Although the multi-temperature maps of Figs. \ref{fig1} and 1.1 look quite 
similar, detailed examination reveals that
the inclusion of the 1.3 mm data has not only increased
the spatial resolution, but has also increased the detection significance of 
the lowest-temperature structures. We base our
subsequent discussion on the results obtained using
the full six-wavelength data set ({\it Herschel\/} plus 1.3 mm), which
yield a final spatial resolution of approximately $8''$.

Figs. \ref{fig1} and 1.1 show that
at the lowest temperature ($T_{\rm D}=13.9$ K) the only visible structure is
a $\sim6$ pc tadpole-like feature with a prominent condensation at its head.
In the next temperature steps, other features become visible,
and at $T_{\rm D}=26.6$ K the structure is dominated by material around the 
outside of the cloud---there is a depression in the interior representing a
deficit of material at that temperature. The deficit becomes even more
pronounced at $T_{\rm D}=31.3$ K, whereby the interior appears to be 
devoid of material. Such behaviour is not unexpected for the Brick itself
since the material at that temperature is believed to be limited to a thin outer
shell \citep{clark13}. The apparent limb brightening
in the $T_{\rm D}=31.3$ K panel is consistent with that prediction.
However, one would nevertheless expect to see some overlying ISM at the Brick 
location, and the assocated uncertainty map for $T_{\rm D}=31.3$ K allows
for the presence of a significant amount of material there, albeit 
below the detectability level.
Moving to still higher temperatures we find that at $T_{\rm D}=36.8$ K,
a new structure emerges in the interior in the form of a thin hot strip 
which persists
up to $T_{\rm D}=51.0$ K. 

\begin{figure}
\includegraphics[width=90mm]{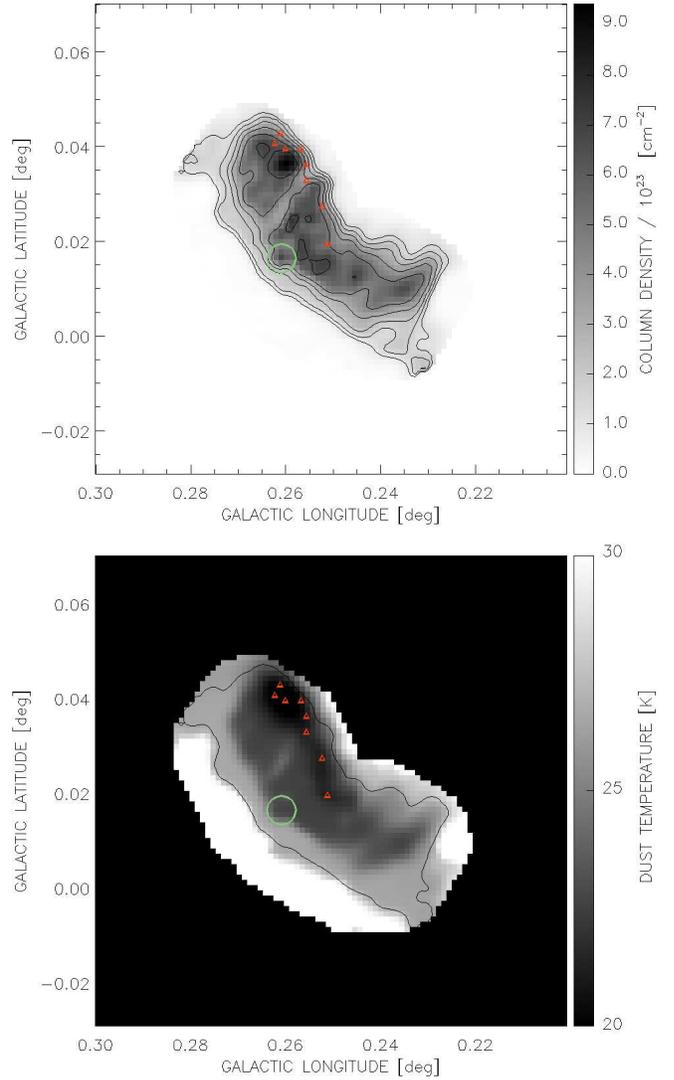}
\caption{Maps of integrated column density and mean temperature.
{\it Upper panel:\/}
Integrated column density, summed over all
temperature planes of the image cube of differential column density.
Contours are plotted in decreasing intervals of $\sqrt{2}$ from the peak
down to a lowest contour level of $1.6\times10^{23}\,{\rm cm}^{-2}$ which
we define as the outer boundary of the Brick.
{\it Lower panel:\/}
Mean (density-weighted) line-of-sight dust temperature. The black curve
represents the outer boundary of the Brick as defined above.
The field of view of both images is 
truncated at the bean-shaped outer boundary of
the maps published by \citet{john14}, corresponding to the edges of the 
SMA primary beam in their 6-pointing mosaic.
On both panels the red triangles represent the locations of local 
maxima of hot dust, as listed in Table 1. Also overplotted (green circle)
is the location of the water maser reported by
\citet{lis94}.}
\label{fig2}
\end{figure}

Fig. \ref{fig2} shows the map of integrated column density obtained by
summing the set of differential column densities from all 12 temperature planes
of the image cube produced by {\tt PPMAP} using the full data set 
({\it Herschel\/} plus 1.3 mm). Also shown in this figure is the corresponding
map of mean line-of-sight dust temperature obtained from a density-weighted
average of the temperatures in the image cube underlying each spatial pixel.
Overplotted on both panels are the locations of 
local density maxima of the hot dust. 
They correspond to the five strongest maxima in the
$T_{\rm D}=43.3$ K map, and all three maxima at $T_{\rm D}=51.0$ K.
They are listed in Table 1 together with
the corresponding values of differential column density
and density-weighted mean temperature for the temperature range, 40--55 K,
which encompasses both of those temperature bins.
The column density and temperature maps in Fig. \ref{fig2} are consistent 
with their counterparts presented by \citet{long12}, taking into account
the differing values of $\beta$ used and the difference in spatial 
resolution\footnote{Repeating our analysis using the \citet{long12} 
value of opacity index,
$\beta=1.75$, gives a total mass of $1.3 \times 10^5\,M_\odot$ as compared to
the Longmore et al.\ value of $1.7 \times 10^5\,M_\odot$. Our peak
column density is then a factor of 2.4 higher than the Longmore et al.\ value
and our minimum dust temperature (15.7 K) is about 3 K lower than theirs.
Tests with synthetic data \citep{mar15} indicate that both
of these apparent discrepancies can be accounted for by the increased
resolution provided by the deconvolution implicit in {\tt PPMAP}. 
We also find that the key morphological
features of our image cube (in particular, the presence of the thin hot strip)
are preserved with the alternate choice of $\beta$, although the
estimated temperatures are lower---the hot
strip then occupies the temperature range 31--43 K rather than
37--51 K.}. 

\begin{table}
 \begin{minipage}{200mm}
  \caption{Local maxima in the high-temperature dust distribution.} 
  \begin{tabular}{@{}ccccc@{}}
  \hline
 Peak\#\footnote{Peaks 1--5 and 6--8 are from the 
43.3 K and 51.0 K maps, resp.} 
 & Gal. long. & Gal. lat.  &  Col. dens.\footnote{Total column density in temperature range $T_{\rm D}=40$--55 K.} / $10^{20}$ & $\overline{T_{\rm D}}$\footnote{Density-weighted mean temperature in that same range.}\\
 &   $[^\circ]$ & $[^\circ]$ &  $[$cm$^{-2}]$  \\
 \hline
      1    &      0.2512   &     0.0197    &    102    &    43.3  \\
      2    &      0.2523   &     0.0275    &    100    &    43.3  \\
      3    &      0.2557   &     0.0331    &     77    &    43.4  \\
      4    &      0.2601   &     0.0397    &     76    &    44.0  \\
      5    &      0.2623   &     0.0408    &     68    &    43.5  \\
      6    &      0.2557   &     0.0364    &     63    &    48.3  \\
      7    &      0.2612   &     0.0431    &     47    &    49.7  \\
      8    &      0.2568   &     0.0397    &     47    &    50.0  \\
\hline
\end{tabular}
\end{minipage}
\end{table}

\begin{figure}
\begin{center}
\includegraphics[width=90mm]{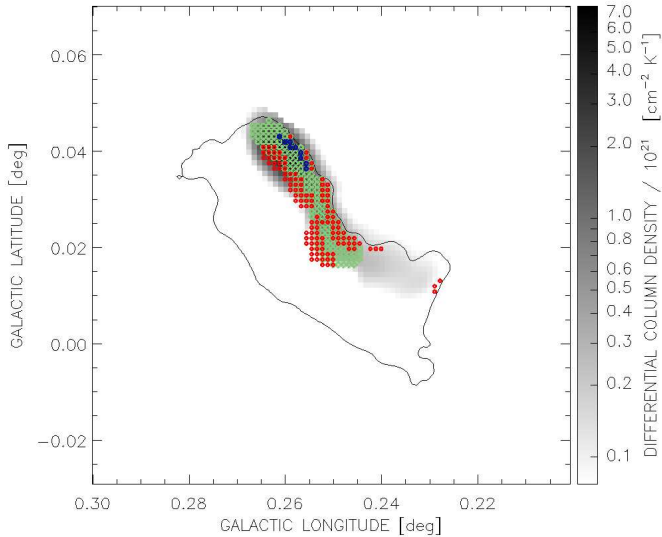}
\caption{The differential column density image at $T=13.9$ K, shown in
greyscale, overlaid by coloured symbols which represent
the locations of hot dust at three additional
temperatures, namely 36.8 K (red), 43.3 K (green), and 51.0 K (blue).
Each plotted symbol (open circle) represents a pixel in which the
differential column density
exceeds the $3\sigma$ level at the particular temperature. 
For comparison, the black curve represents the outer boundary of the Brick
as defined in Fig. \ref{fig2}.}
\label{fig3}
\end{center}
\end{figure}

Interestingly, the thin hot strip visible in the temperature range 36.8--51.0 K
closely overlays the much cooler tadpole feature at 13.9 K. This is
illustrated by Fig. \ref{fig3} which shows a greyscale representation of the
differential column density image of 13.9 K material, overplotted with
coloured symbols representing the set of all pixels in the
36.8 K, 43.3 K, and 51.0 K images which exceed the $3\sigma$ uncertainty
level. This figure shows that, in projection, the structure at all four 
temperatures is closely coincident, with the
higher-temperature material nested within the lower-temperature material.
This behaviour suggests a cylindrical
structure in which the temperature increases outwards.

\begin{figure}
\includegraphics[width=84mm]{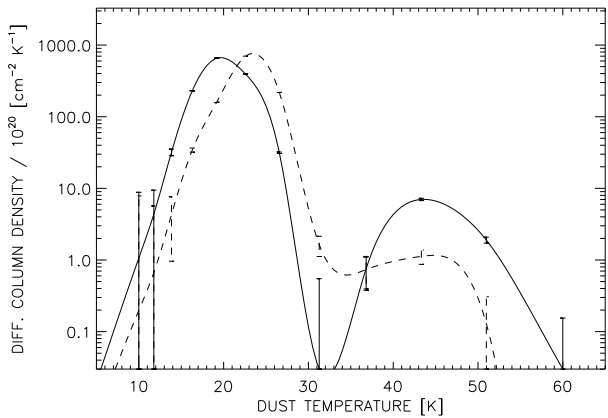}
\caption{Profiles of differential column density as a function of dust 
temperature. The solid curve represents the average over
eight separate lines of sight corresponding to the locations listed in Table 1
and plotted as red triangles in Fig. \ref{fig2}. 
For comparison, the dashed curve shows the mean profile for the Brick, 
spatially averaged over the region within the Brick boundary defined 
in Fig. \ref{fig2}.}
\label{fig4}
\end{figure}

The mean temperature profile through the ``tadpole"
feature (averaged over eight lines of sight, each of which corresponds to
a pixel located at a local maximum as listed in Table 1 and represented
by a red triangle in Fig. \ref{fig2}) is shown in Fig. \ref{fig4}. 
It is characterised by the presence of
two distinct peaks, indicating that the cool and hot material are
physically separated at some spatial scale.
The same is true of all eight profiles individually.
Also shown, for comparison, is the mean temperature
profile for the Brick itself, averaged over the region within the
boundary defined in Fig. \ref{fig2}.

\section[]{Discussion}

Much of the dust temperature structure that we have mapped is consistent
with external heating of the Brick by the ISRF, as has been discussed
in previous work \citep{lis01,long12,clark13}. In particular, the
warm outer sheath manifest as limb brightening at $T_{\rm D}=31.3$ K is 
consistent with the predictions
of the simulation by \citet{clark13}. However, we have found that within
this cloud, near one edge, is a cool ($T_{\rm D}\sim14$ K) feature with a 
tadpole-like geometry, and that a much hotter structure lies within
it, at least in 2D line-of-sight projection. Furthermore, there is evidence for
temperature stratification such that the hottest material ($T\sim 50$ K)
exists within a thin central spine and that the temperature decreases
outwards down to the value of $T\sim14$ K. If the structure were heated
externally, we would expect to see the hottest material at the very edge, but
Fig. \ref{fig3} shows that this is not the case, i.e., the temperature actually 
decreases at the edge. Such behaviour is inconsistent with
external heating and it is likely, therefore, that the structure is heated
internally. We suggest that it
represents a cool filament whose hot cylindrical spine
is collapsing and fragmenting to form a line of protostars. 
We estimate the mass and luminosity of this spine by
summing the contributions from material at each of the above three 
temperatures, and obtain values of $\sim 800\,M_\odot$ and
$\sim6\times 10^4\,L_\odot$, respectively. The location of peak
dust temperature ($\sim51$ K), in Galactic coordinates, is
$[\ell,b]=[0.\!\!^\circ 2578, 0.\!\!^\circ0397]$. 

If indeed the hot spine represents a locus of star formation, then its
location near the edge of the cloud is
reminiscent, in some ways, of part of the ``wishbone"
structure of Orion A which has been modelled by \citet{hart07}
in terms of nonlinear gravitational acceleration effects. 
In such models, a star-forming condensation arises at one
end of a sheet-like molecular cloud, with density enhancements extending 
along the two edges. The fact that the \citet{lis94} water maser 
is also located near an edge, on the opposite side of the Brick, would
be consistent with such a scenario. In any event, further observations at
high spatial resolution will be necessary to confirm the presence of
star formation in the hot spine.

\section*{Acknowledgments}

We thank K. Johnston for a helpful discussion, and also thank the referee
for helpful comments.
This research is supported by the EU-funded {\sc vialactea} Network (Ref. FP7-SPACE-607380).

\bsp

\label{lastpage}


\begin{thebibliography}{99}
\bibitem[\protect\citeauthoryear{Clark et al.}{2013}]{clark13} 
  Clark, P. C., Glover, S. C. O., Ragan, S. E. et al. 2013, ApJ, 768, L34
\bibitem[\protect\citeauthoryear{Ginsburg et al.}{2016}]{gin16} 
  Ginsburg, A., Henkel, C., Yipling, A. et al. 2016, A\&A, 586, 50
\bibitem[\protect\citeauthoryear{Griffin et al.}{2013}]{griffin13} 
  Griffin, M. J., North, C. E., Amaral-Rogers, A. et al. 2013, MNRAS, 434, 992
\bibitem[\protect\citeauthoryear{G{\"u}sten et al.}{1981}]{guesten81} 
G{\"u}sten, R., Walmsley, C.~M., \& Pauls, T.\ 1981, A\&A, 103, 197 
\bibitem[\protect\citeauthoryear{Hartmann \& Burkert}{2007}]{hart07} 
  Hartmann, L. \& Burkert, A. 2007, ApJ, 654, 988
\bibitem[\protect\citeauthoryear{Johnston et al.}{2014}]{john14} 
  Johnston, K. G., Beuther, H., Linz, H. et al. 2014, A\&A 568, A56
\bibitem[\protect\citeauthoryear{Kauffmann et al.}{2013}]{kauff13} 
  Kauffmann, J., Pillai, T. \& Zhang, Q. 2013, ApJ, 765, L35
\bibitem[\protect\citeauthoryear{Kruijssen et al.}{2015}]{kruij15} 
Kruijssen, J.~M.~D., Dale, J.~E., \& Longmore, S.~N.\ 2015, MNRAS, 447, 1059 
\bibitem[\protect\citeauthoryear{Lis et al.}{1994}]{lis94} 
  Lis, D. C., Menten, K. M., Serabyn, E. \& Zylka, R. 1994, ApJ, 423, L39
\bibitem[\protect\citeauthoryear{Lis et al.}{2001}]{lis01} 
  Lis, D. C., Serabyn, E., Zylka, R., \& Li, Y. 2001, ApJ, 550, 761
\bibitem[\protect\citeauthoryear{Longmore et al.}{2012}]{long12} 
  Longmore, S. N., Rathborne, J., Bastian, N. et al. 2012, ApJ, 746, 117
\bibitem[\protect\citeauthoryear{Longmore et al.}{2013}]{long13} 
Longmore, S.~N., Kruijssen, J.~M.~D., Bally, J., et al.\ 2013, MNRAS, 433, L15 
\bibitem[\protect\citeauthoryear{Marsh et al.}{2015}]{mar15} 
  Marsh, K. A., Whitworth, A. P. \& Lomax, O., MNRAS, 454, 4282
\bibitem[\protect\citeauthoryear{Mills \& Morris}{2013}]{mills13} 
Mills, E.~A.~C., \& Morris, M.~R.\ 2013, ApJ, 772, 105 
\bibitem[\protect\citeauthoryear{Mills et al.}{2015}]{mills15} 
  Mills, E. A. C., Butterfield, N., Ludovici, D. A. et al. 2015, ApJ, 805, 72
\bibitem[\protect\citeauthoryear{Molinari et al.}{2010}]{mol2010} 
  Molinari, S., Swinyard, B., Bally, J., et al. 2010, PASP, 122, 314
\bibitem[\protect\citeauthoryear{Molinari et al.}{2011}]{mol2011} 
Molinari, S., Bally, J., Noriega-Crespo, A., et al.\ 2011, ApJ, 735, L33 
\bibitem[\protect\citeauthoryear{Ossenkopf \& Henning}{1994}]{oss94} 
  Ossenkopf, V. \& Henning, T. 1994, A\&A, 291, 943
\bibitem[\protect\citeauthoryear{Pezzuto}{2013}]{pez13} 
  Pezzuto, S. 2013, HERSCHEL Document Ref.: PICC-CR-TN-044
\bibitem[\protect\citeauthoryear{Pilbratt et al.}{2010}]{pilb10} Pilbratt,
G. L., Riedinger, J. R., Passvogel, T., et al. 2010, A\&A, 518, L1
\bibitem[\protect\citeauthoryear{Poglitsch et al.}{2010}]{pog10} Poglitsch, A.,
  Waelkens, C., Geis, N., et al. 2010, A\&A, 518, L2
\bibitem[\protect\citeauthoryear{Rathborne et al.}{2014a}]{rath14} Rathborne,
  J. M., Longmore, S. N., Jackson, J. M. et al. 2014a, ApJ, 786, 140
\bibitem[\protect\citeauthoryear{Rathborne et al.}{2014b}]{rath14b} Rathborne,
  J. M., Longmore, S. N., Jackson, J. M. et al. 2014b, ApJ, 795, L25
\bibitem[\protect\citeauthoryear{Rathborne et al.}{2015}]{rath15} Rathborne,
  J. M., Longmore, S. N., Jackson, J. M. et al. 2015, ApJ, 802, 125
\bibitem[\protect\citeauthoryear{Valtchanov}{2014}]{valt14} 
  Valtchanov, I. 2014, HERSCHEL-DOC-0798, version 2.5, March 24, 2014
\end{thebibliography}
\end{document}